\documentclass[amssymb,amsmath,floatfix,superscriptaddress]{revtex4}
\usepackage{graphicx}
\usepackage{amssymb}
\newcommand{\ket}[1]{|#1 \rangle}

\begin{document}

\title{Architectural design for a topological cluster state quantum computer}
    \author{Simon J. Devitt}
    \email{devitt@nii.ac.jp}
    \affiliation{National Institute for Informatics, 2-1-2 Hitotsubashi, Chiyoda-ku, Tokyo 101-8430, Japan.}
     \author{Austin G. Fowler}
    \address{Institute for Quantum Computing, University of Waterloo, Waterloo, Canada.}
 \author{Ashley M. Stephens}
    \affiliation{Centre for Quantum Computer Technology, School of Physics,
    University of Melbourne, Victoria 3010, Australia}
      \author{Andrew D. Greentree}
    \affiliation{Centre for Quantum Computer Technology, School of Physics,
    University of Melbourne, Victoria 3010, Australia}
        \author{Lloyd C.L. Hollenberg}
    \affiliation{Centre for Quantum Computer Technology, School of Physics,
    University of Melbourne, Victoria 3010, Australia}
      \author{William J. Munro}
    \affiliation{Hewlett-Packard Laboratories, Filton Road, Stoke Gifford, Bristol BS34 8QZ, United Kingdom}
      \affiliation{National Institute for Informatics, 2-1-2 Hitotsubashi, Chiyoda-ku, Tokyo 101-8430, Japan.}
        \author{Kae Nemoto}
    \affiliation{National Institute for Informatics, 2-1-2 Hitotsubashi, Chiyoda-ku, Tokyo 101-8430, Japan.}
\date{\today}

\begin{abstract}
The development of a large scale quantum computer is a highly sought after goal of fundamental research 
and consequently a highly non-trivial problem.  Scalability in quantum information processing is not just a 
problem of qubit manufacturing and control but it crucially depends on the ability to adapt advanced techniques in
quantum information theory, such as error correction, to the experimental restrictions of 
assembling qubit arrays into the millions. 
In this paper we introduce a feasible architectural design for large scale quantum computation in optical systems.  
We combine the recent developments in topological cluster state computation with the photonic module, a simple 
chip based device which can be used as a fundamental building block for a large scale computer.  The integration of 
the topological cluster model with this comparatively simple operational element addresses many significant issues 
in scalable computing and leads to a promising modular architecture with complete integration of 
active error correction exhibiting high fault-tolerant thresholds.  
\end{abstract}

\maketitle
\section{Introduction}
The scientific effort to construct a large scale device capable of
Quantum Information Processing (QIP) has advanced significantly over the past 
decade~\cite{NPT99,CNHM03,YPANT03,CLW04,HHB05,GHW05,G06,H06,G07,O'Brien2}.  
However, the design of a viable processing architecture capable of housing and manipulating the 
millions of physical qubits necessary for scalable QIP is hampered by an architectural design 
gap that exists between advanced 
techniques in error correction and fault-tolerant computing and the 
comparatively small scale devices under consideration by experimentalists.  
While some progress in scalable system design has been made~\cite{Kielpinski1,Taylor1,Hollenberg1,Fowler3,Knill1,Kok1}, 
the development of a truly large scale quantum architecture, able to implement programmable 
QIP with extensive error correction, hinges on the ability to 
adapt well established computational and error correction models to the 
experimental operating conditions and fabrication restrictions 
of physical systems, well beyond 1000 physical qubits.  

The recent introduction by Raussendorf, Harrington and Goyal of
Topological Cluster state Quantum
Computation (TCQC)~\cite{Raussendorf4,Raussendorf5,Fowler2} marks an
important milestone in the advance of theoretical techniques for
quantum information processing (QIP).  This model of QIP differs
significantly from both the circuit model of computation and the
traditional cluster state model~\cite{NC00,Raussendorf1} in that the
entanglement resource defines a topological ``backdrop" where
information qubits can be defined in very non-local ways. By
disconnecting the physical qubits (used to construct the lattice)
from the information qubits (non-local correlations within the
lattice), information can be topologically protected, which 
is inherently robust against local perturbations caused
by errors on the physical system. This exploitation of
topological protection utilizing qubits, rather than exotic anyonic
systems, which are particularly difficult to experimentally
create and manipulate~\cite{Nayak1}, leads to a 
computation model that exhibits high fault-tolerant thresholds where
problematic error channels such as qubit loss are naturally
corrected.

In terms of architecturally implementing certain topological coding models, there are two general 
techniques which can be used, depending on the physical system under consideration.  
Surface codes~\cite{BK01,DKLP02,Fowler1}, appropriate for matter based 
systems and the 3D TCQC model~\cite{Raussendorf4,Raussendorf5}, which is far more useful 
in optics based architectures 
due to the high mobility and comparatively inexpensive cost of photonic qubits.
This paper introduces a computational architecture for the 3D
topological model, utilizing photonic qubits.  By making use of the TCQC model, 
error correction protocols are automatically incorporated as a property of the computational 
scheme and photon loss (arguably the dominant error channel) becomes 
correctable without the incorporation of additional coding schemes or 
protocols~\cite{Dawson2,Ralph1,Varnava1,Kok1}.

As error correction within the TCQC model is predicated on the presence of a large
cluster lattice before logical qubits are defined, standard probabilistic techniques for 
preparing entangled photons are generally not appropriate without employing extensive 
photon routing and/or storage.  Therefore, a deterministic photon/photon entangling gate 
is desired to drastically simplify any large scale implementation of this model.  Such operations 
could for instance include the C-Phase and parity gates~\cite{Beenakker4,Nemoto4,Spiller6}, 
and may be via direct photon-photon interactions, cavity QED
techniques~\cite{Domokos1,Duan1} or an indirect bus-mediated quantum
nondemolition-like interaction~\cite{Spiller6,Devitt1}. As an architectural building block, 
we are going to focus on the later with the recently introduced idea of the
photonic module and chip~\cite{Devitt1,Stephens1}.
This small scale, chip based quantum device is an illustrative example of a technology
with all the essential components to act as an architectural building block for the TCQC model. 
This device has the flexibility to entangle an arbitrary number of photons with no
dynamic change in its operation and the manner in which photons
flow in and out of the unit make it an ideal component to realize
a modular structure for large scale TCQC.
In recent years, experimental work in the both cavity/photon 
coupling~\cite{E1,F1,F2} and chip based single photon technologies~\cite{P1,C1,M1} 
has advanced significantly.  While the high fidelity construction of the photonic chip is 
still a daunting task, the continued effort in these areas allows for optimism that such a 
device is an experimental possibility. 

\begin{figure*}[ht!]
\begin{center}
\resizebox{130mm}{!}{\includegraphics{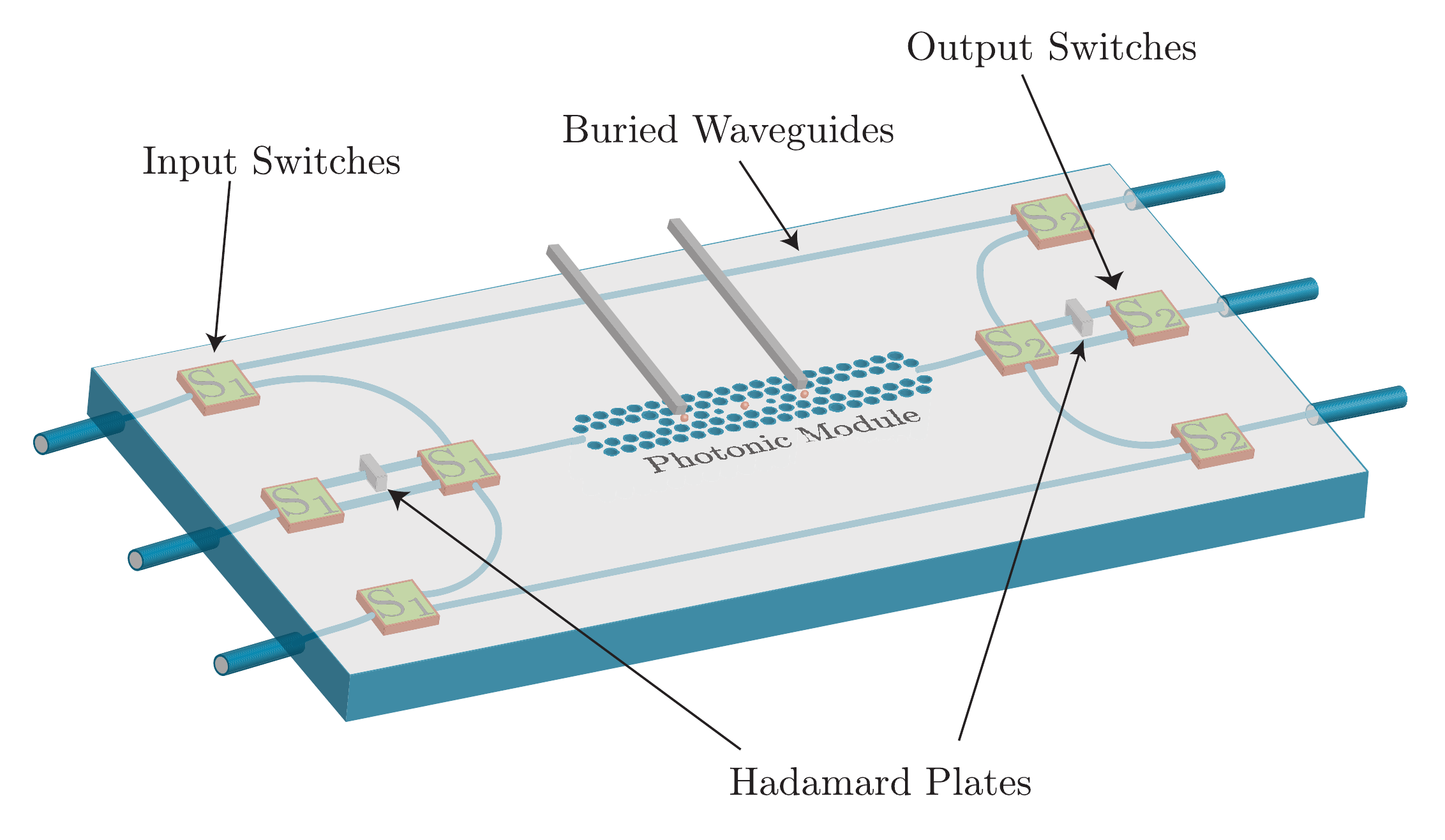}}
\end{center}
\vspace*{-10pt}
\caption{\textbf{Schematic design for the photonic chip}.  The chip is a 3-in 3-out
integrated device containing one photonic module~\cite{Devitt1}, classical single photon routing 
and two optical waveplates allowing for the optional application of single photon 
Hadamard gates to specific photons.  Not shown are Hadamard
plates potentially required on the input or output port of the chip.}
\label{figure:chip}
\end{figure*}

The following will detail the lattice preparation
network, utilizing the photonic chip illustrated in
Fig.~\ref{figure:chip} as the basic building block of the computer.
We detail the physical layout of the network, the optical
switching sequence required and several techniques that are used to
optimize the preparation of the cluster.  We complete the discussion
with a resource analysis, examining the number of fabricated devices
required to perform a large scale quantum algorithm.

As this paper is focused on the possible construction of a large scale 
architecture, we utilize the photonic chip as a building block for this architectural design and we direct readers to 
Refs.~\cite{Devitt1,Stephens1,Su1,Yang1} which examine the micro-analytics of the photonic chip in more depth.
Additionally, we will 
also assume (for the sake of conceptual simplicity) 
that appropriate, high fidelity, on demand single photon 
sources and detectors are also feasible on the same developmental time frame as a photonic chip.  
However, as the photonic module is essentially a non-demolition photon detector, sources 
and detectors can be constructed directly with the chip~\cite{Nemoto1} allowing us to relax 
these assumptions, if required. 

\section{The flowing computer design}
\begin{figure*}
\begin{center}
\resizebox{130mm}{!}{\includegraphics{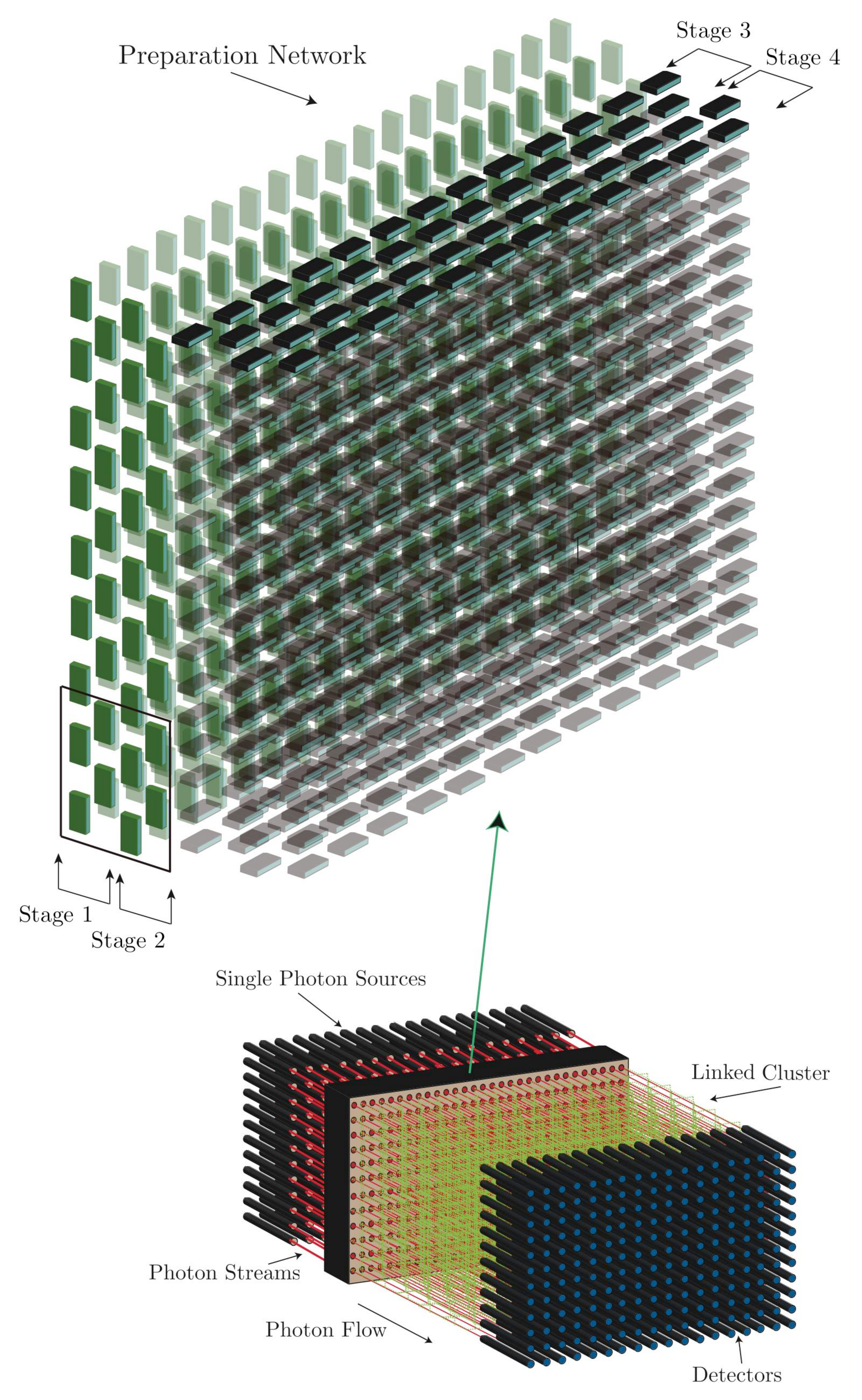}}
\end{center}
\vspace*{-10pt}
\caption{\textbf{General layout of the optical architecture}.  
Layers of photonic chips are arranged in 4 stages to prepare the entanglement
links between qubits.  Photons enter the network from appropriate single photon sources at the 
rear and exit the front
linked up in the required cluster state.  
Single photon detectors can then be placed immediately after the preparation network to consume the cluster, performing
computation.  This eliminates the need for photon routing and storage as each individual photon essentially follows a linear trajectory from the source, through the
photonic chip preparation network and into detectors.  The box in the left hand figure 
indicates the small section of stages 1 and 2 of the preparation network 
illustrated in Fig.~\ref{figure:side}.}
\label{figure:network}
\end{figure*}

The general structure of the optical TCQC is illustrated in Fig.~\ref{figure:network}.  
The high mobility of 
single photons and the nature of the TCQC model allows for an essentially ``flowing" 
model of computation.  Initially,
unentangled photons enter the preparation network from the rear,
flow through a static network consisting of layers of photonic chips arranged in 4 stages and exit the
preparation network where they can be immediately consumed for
computation.
\begin{figure*}[ht!]
\begin{center}
\resizebox{130mm}{!}{\includegraphics{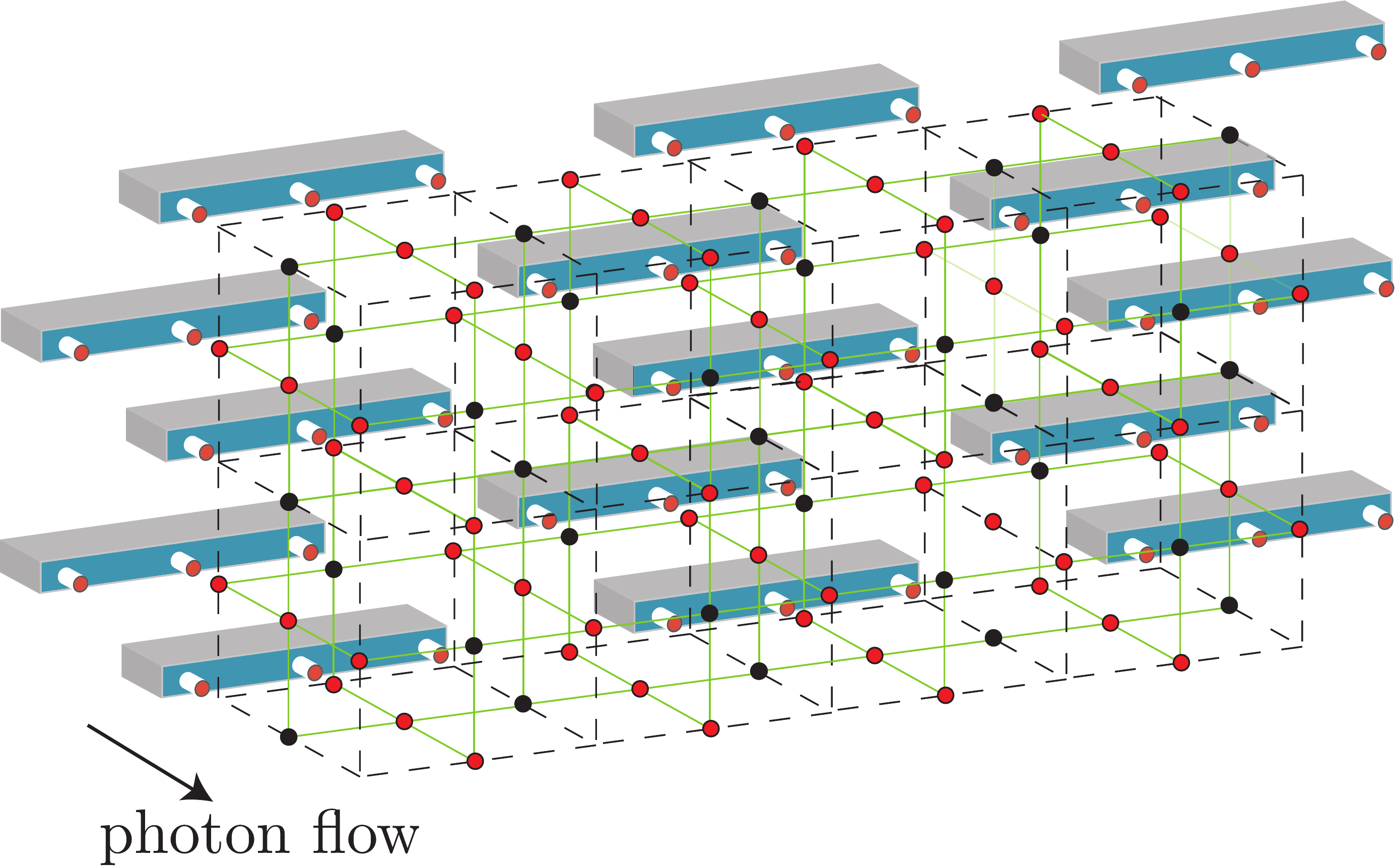}}
\end{center}
\vspace*{-10pt}
\caption{\textbf{Layout of the optical architecture}.
This is a magnified section of Fig.~\ref{figure:network} where only the last (of four) stages 
of photonic chips is illustrated.}
\label{figure:cluster}
\end{figure*}
\begin{figure}[ht!]
\resizebox{70mm}{!}{\includegraphics{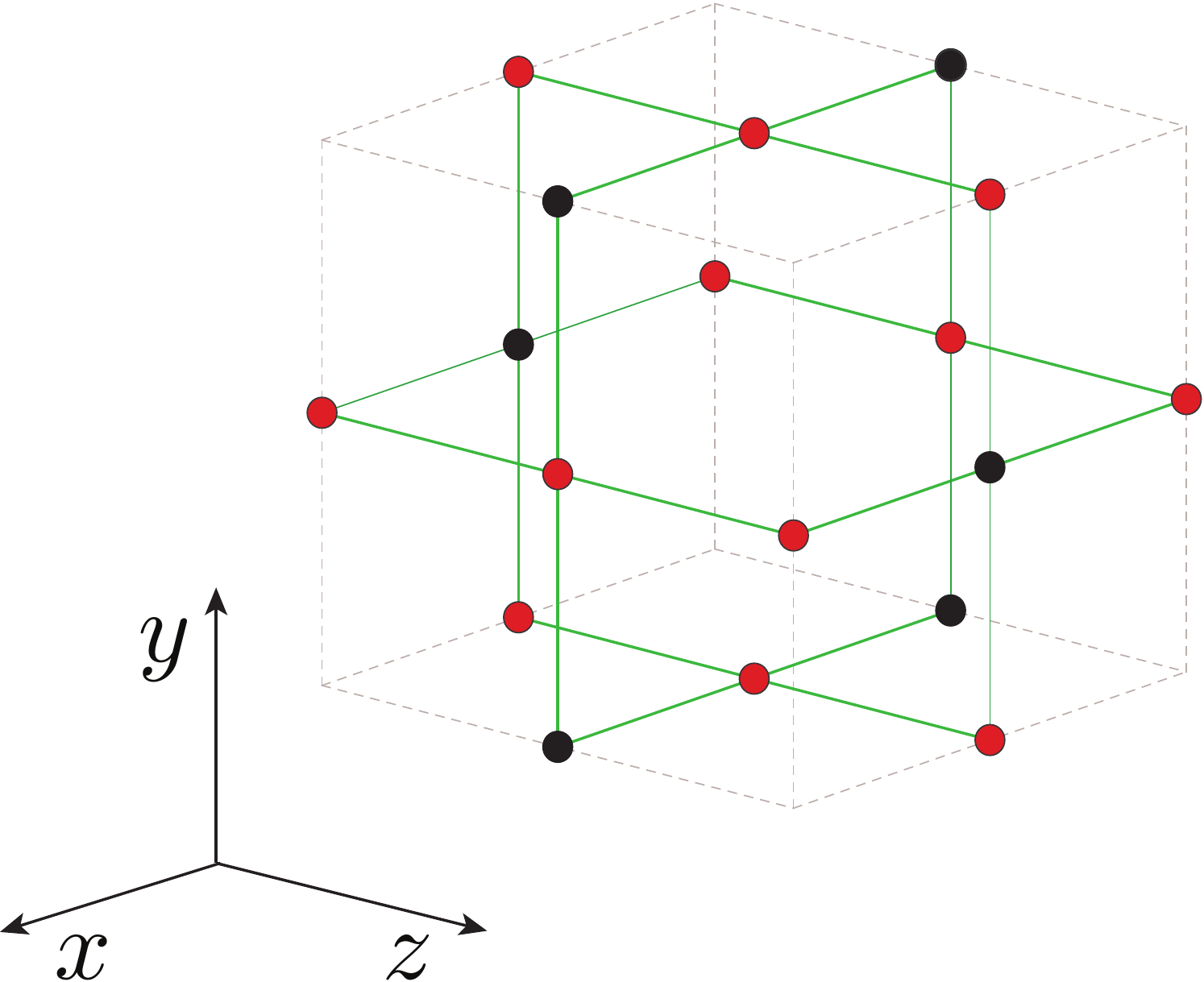}}
\caption{\textbf{Structure of each unit cell of the cluster}.  The cluster is not a Body Centered Cubic (BCC) 
lattice.  The green links, representing entanglement bonds specified by the stabilizer 
generators of Eq.~\ref{eq:stab}, ensure that each photon is 
connected to four neighbors, 
not six.  In the flowing network, photons along the $z$-axis (for a given $(x,y)$ co-ordinate) 
are carried by common optical waveguides.  As the lattice is
not fully connected, there are two groups of waveguides, 
one set runs at a fixed repetition rate along the $z$-axis (the red photons shown above) and the other
operates at half that rate (black photons).  We denote these waveguides as 
full and half rate lines respectively.}
\label{figure:unit}
\end{figure}
Fig~\ref{figure:cluster}. shows a smaller region of
computer where only one (of four) stages of the
preparation network is illustrated for clarity while Fig~\ref{figure:unit}. 
illustrates a single unit cell of the cluster state which repeats and extends in all three
dimensions. In the TCQC model, computation proceeds by measuring, 
sequentially, cross sectional layers ($x-y$ plane) of the 
lattice which acts to simulate time through the computation.  Qubit information is defined within the lattice 
via the creation of holes (known as defects).  As the cluster is consumed along the $z$-axis, defects 
are deformed via measurement such that they can be braided around each other, enacting CNOT 
gates between qubits~\cite{Raussendorf5,Fowler2}.

Each unit cell of the cluster does not have a Body Centered Cubic (BCC) structure, each photon is only connected to 
four neighbors, not six.  Therefore, in a flowing network design, where each photon pulse along the 
$z$-axis is carried by common optical waveguides, there are two sets of photon pulse repetition rates.  
One set (waveguides carrying red photons in Fig~\ref{figure:unit}) runs at a fixed repetition 
rate which we will denote as
full rate lines.  The other set (waveguides carrying black photons in Fig~\ref{figure:unit}) 
runs at half that rate, denoted half rate lines. If the lattice is examined along the 
$z$-axis, these full and half rate lines form a checkerboard
pattern.

The global cluster is uniquely defined utilizing the stabilizer formalism~\cite{Gottesman1}. 
The lattice is a unique state defined as the simultaneous $+ 1$ eigenstate 
of the operators,
\begin{equation}
K^{x,y,z} = X_{x,y,z}\bigotimes_{b \in n(x,y,z)} Z_b, 
\label{eq:stab}
\end{equation}
where $X$ and $Z$ are the $2\times 2$ Pauli
operators, $(x,y,z)$ are the co-ordinates of each of the $N$ photons
in the 3D lattice, $n(x,y,z)$ is the set of 4 qubits linked to the
node $(x,y,z)$ and the $N-5$ identity operators are implied [Fig.~\ref{figure:unit}]. For
photons that do not have four nearest neighbors (i.e. at edges of
the lattice), the associated stabilizer retains this form but
excludes the operator(s) associated with the missing neighbor(s).

The optical network to prepare such a state requires the deterministic projection 
of a group of unentangled photons into the
entangled state defined via the stabilizers.  Fig~\ref{figure:chip} illustrates
the basic structure of the photonic chip, introduced in Refs.~\cite{Devitt1,Stephens1} which 
we utilize to perform these deterministic projections.
The photonic module, which lies at the heart of each chip, 
is designed to project an arbitrary $N$ photon state into a $\pm 1$ eigenstate of the
operator $X^{\otimes N}$~\cite{Devitt1}, where
$N$ is the number of photons sent through the module 
between initialization and measurement of the atomic system.  

In order to perform a parity check of the operator $ZZXZZ$, single photon Hadamard 
operations are applied to every photon before and after passage through a photonic chip.  
For every group of five photons sent through 
between initialization and measurement, the central
photon in the stabilizer operator will be routed through a second set of Hadamard 
gates before and after passing through the module.  Note, as each chip is linked in series, the 
Hadamard rotations on each input/output are not required for all stages of the preparation network as they cancel.

\subsection*{Photon stream initialization}
In total there are four stages required to prepare the 3D lattice.
The four stages are partitioned into two groups which have identical layout and switching patterns.  
These two groups stabilize
the lattice with respect to the stabilizer operators along the $y-z$ planes of the cluster 
and the $x-z$ planes respectively.
If an arbitrary input state is utilized to prepare the cluster, two additional stages are 
required to stabilize the cluster with respect
to each operator associated with the $x-y$ plane of each unit cell.  
However we can eliminate the need to perform these parity checks by carefully 
choosing the initial product state that is fed into the preparation network.

Each stabilizer operator, $ZZXZZ$, in the $x-y$ plane has the 
$X$ operator  centered on the photons in half rate lines (i.e.
the black photons in Fig.~\ref{figure:cluster}).  
Each photon in these lines is prepared in the 
$+1$ eigenstate of $X$, $\ket{+} = (\ket{H}+\ket{V})/\sqrt{2}$, 
while photons in full rate lines are
initialized in the state $+1$ eigenstate of $Z$, $\ket{H}$, (we assume 
a polarization basis for computation).  This
initialization ensures that the stabilizer set for the photon stream (before entering the 
preparation network) is 
described by the stabilizers,
\begin{equation}
K^i = X_i, \ i\in {h}, \quad K^j = Z_j, \ j\in {f},
\label{eq:init}
\end{equation}
where $h$ and $f$ are
the sets containing photons in the half and full rate optical waveguides
respectively.  As any product of stabilizers is also a stabilizer of
the system, all of the stabilizers in the $x-y$ planes of the cluster are
automatically satisfied.  It can be easily checked that the only element 
in the group generated by Eq.~\ref{eq:init} that commutes with the stabilizer 
projections associated with the $y-z$ and $x-z$ planes of the cluster 
is the $ZZXZZ$ term associated with each stabilizer along the $x-y$ plane.

\section{Preparation network and photonic chip operation}
\begin{figure*}[ht]
\resizebox{180mm}{!}{\includegraphics{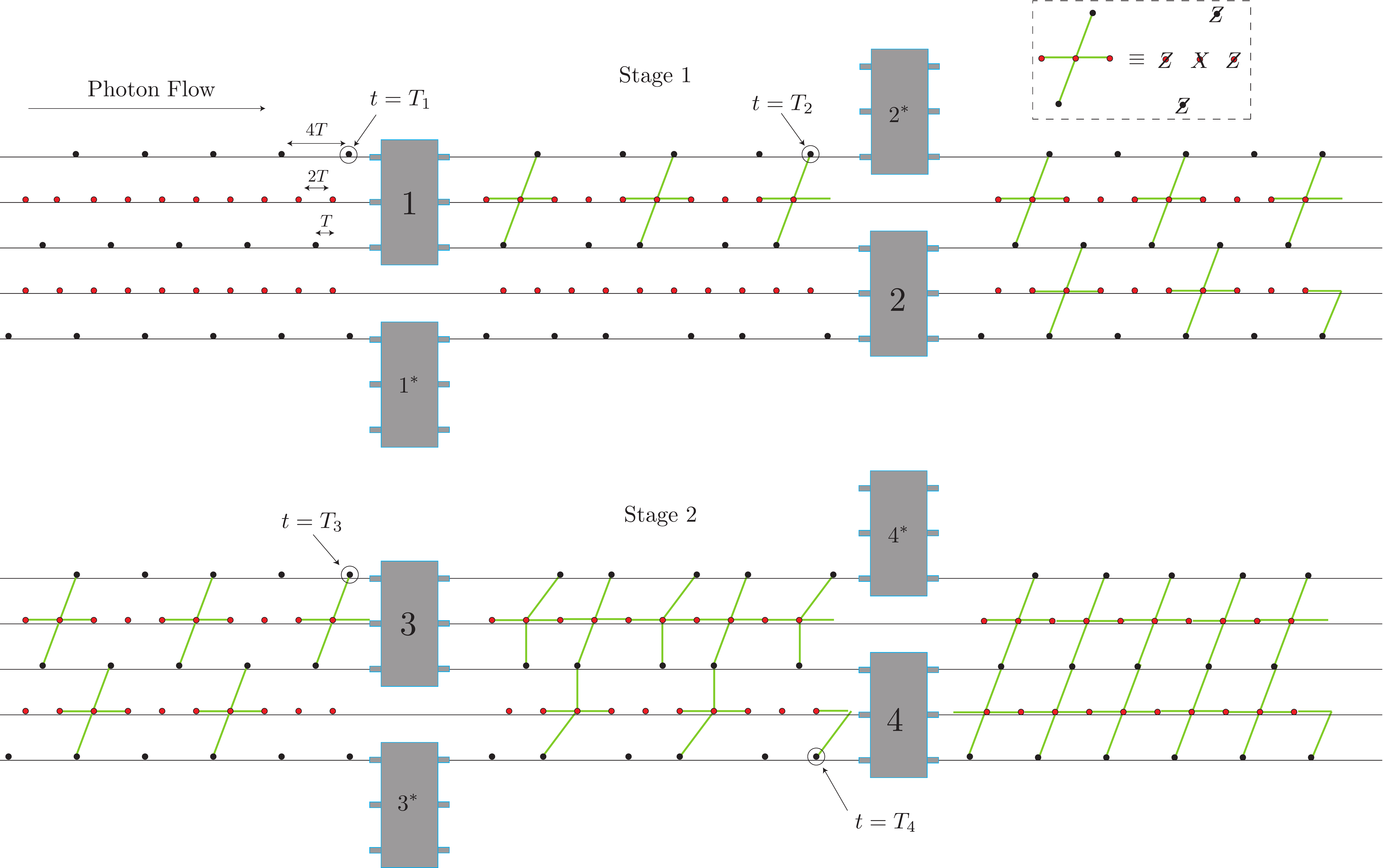}}
\vspace*{-10pt}
\caption{\textbf{Chip orientation for stages one and two of the preparation network, utilized to measure the stabilizer operators along the $y-z$ plane of the lattice}.  
Photons flow from left to right through each chip.  The temporal staggering of each photon is
required such that only one photon passes through each photonic module at any given time step.  The fundamental
temporal staggering is the atom/photon interaction time $T$, however each row has to be temporally offset to ensure
proper temporal ordering for future stages.  Each stage measures a specific stabilizer, illustrated with a cross after the
specific stage.  The operator that is measured is $ZZXZZ$, where the $X$ operator associated with the photon at the center of the
cross (insert).  Extending this network to more and more cells requires extending each column vertically.  The network for stages three and
four, required to stabilize the system with respect to the $x-z$ plane, is identical where 
the switching sequence is offset by $2T$.}
\label{figure:side}
\end{figure*}

Illustrated in Fig~\ref{figure:side} is a 2D cross section of the preparation network 
(the section of the global preparation network indicated in Fig.~\ref{figure:network}).  
We illustrate only two stages of the
network since the preparation network in the $x-z$ plane is identical to the $y-z$ plane.
Each stage requires a staggered arrangement of
photonic chips and in Fig~\ref{figure:side}, after a specific stage,
we have detailed the stabilizer that has been measured where the central
photon corresponds to the $X$ term in the measured operator.  The temporal staggering of the
photonic state, for each optical line, is also detailed where the spacing interval, $T$, 
is bounded below by the minimum interaction
time required for the operation of the photonic module and 
bounded above by the coherence time of the atomic systems in
each chip~\cite{Devitt1,Stephens1}.

\subsection{Temporal Asynchrony} 
Before detailing the switching sequence for these stages we have to address a slight complication that arises when creating the 3D lattice.
In general, each photon is involved in five separate parity checks.
By utilizing a specific input state we can remove the need to measure the stabilizers 
associated with the $x-y$ planes of each unit cell.  This implies that
we are measuring four operators for each photon present in half rate lines 
(we no longer perform a parity check associated with the
$x-y$ plane stabilizers for these photons) and three operators 
for each photon in full rate lines (each
of these photons are involved in two parity checks associated with $x-y$ plane stabilizers).  As each photon suffers a temporal delay of $T$ by passing through
a photonic module, there is a temporal discrepancy with the photons in the half rate and full rate 
lines which, if not compensated, leads to two photons
being temporally synchronous in later parity checks.
The flexibility of the photonic chips allows us to solve the 
delay problem in a convenient way.

If a delay were not required, a given module would have a window of $3T$ where it is idle 
between successive parity checks.  Normally this window would be utilized for 
measurement and re-initialization of the atomic system.  
For a given stage of the preparation network, every fourth photon in the full
rate lines is not involved in any parity check (see Fig.~\ref{figure:side}) and must 
therefore be delayed.  Hence we partition this $3T$ window into three steps.
Immediately after the last photon from the previous parity check exits the chip, the atomic system is measured (now utilizing a
temporal window of $T$ rather than $3T$).  As the delayed photon is the next one to enter the
chip, we simply do not initialise the module in the required atomic superposition state.  

As detailed in Ref.~\cite{Stephens1}, the atom/photon interaction
required for the module requires initializing a 3-level atomic system into a 
superstition of its two ground states, $(\ket{1}+\ket{2})/\sqrt{2}$, with a resonant RF field.
The presence of a photon in the
cavity mode (which is detuned with the $\ket{1} \leftrightarrow \ket{3}$ transition) induces a phase shift on the state $\ket{1}$ which oscillates the
system between the states $(\pm \ket{1}+ \ket{2})/\sqrt{2}$.  If we instead keep the 
atomic system in the $\ket{1}$ (or $\ket{2}$) state, the
presence of a photon will have no effect on the atomic system.  
Delaying the required photon by $T$ therefore
requires operating the module as usual, but for this first step we do not 
initialize the atomic system.  This stage again takes time $T$
and we refer to it as the ``holding" stage.
In the next temporal window, $T$, the atomic system is re-initialized in
the appropriate superposition state and the next 5 photon 
parity check proceeds as normal.  

This delay trick will maintain the temporal
staggering of the photons without the need for additional technology
and ensures that every photonic chip is active for all time steps
(no idle time for any module).  

\subsection{Switching sequence}

Illustrated in Fig.~\ref{figure:side} and Tab.~\ref{tab:side} is 
the network and switching sequence for the stabilizers associated with the
$y-z$ planes of the unit cell.  The stabilizers associated with the $x-z$ planes 
are measured using precisely the same network and
switching sequence (when viewing the lattice along the $y$-axis).  
The difference in the switching sequence is a global offset of $2T$
to account for the delay from stages one and two.
\begin{table}[ht!]
\begin{center}
\vspace*{4pt}
\begin{tabular}{c|c|c|c|c|c|c|c|c|c|c|c|c|c}
Chip & 1&2&3&4&5&6&7&8&9&10&11&12&13 \\
\hline
1 & U&C&B&L&M&H&I&R&U&C&B&L&M\\
1* & M&H&I&R&U&C&B&L&M&H&I&R&U\\
2 & B & L & M & $H^U$ &I&$R^B$&U&C&B&L&M&$H^U$&I\\
2* & I & $R^B$ & U & C &B& L &M&$H^U$&I&$R^B$&U&C&B\\
3 & B & L & M &H&I&R&U&C&B&L&M&S&I\\
3* & I & R& U &C&B&L&M&H&I&L&U&C&B\\
4  & M & H$^U$&I&R$^B$&U&C&B&L&M&H$^U$&I&R$^B$ & U\\
4*  & U & C&B&L&M&H$^U$&I&R$^B$&U&C&B&L & M
\end{tabular}
\vspace*{4pt} \caption{\textbf{Switching sequence for stages one and two of
the photonic chip network.}   We have not illustrated stages three and
four as the spatial arrangement of chips and the switching sequences
are identical when examining a cross section the cluster along the $y$-axis
(the switching sequence for stages three and four will have a $2T$
offset from the one shown above).  The temporal location of each of
the photons at $t=\{T_1,T_2,T_3,T_4\}$ is indicated in
Fig.~\ref{figure:side}.  The labels $R,L,C,U,B$ refer to the
geometric arrangement of the photons in the lattice stabilizer (i.e.
right, left, center, upper and bottom) while the switching settings
are $R = C = L = \text{central chip port}$, $U = \text{upper chip
port}$, $B = \text{lower chip port} $, $M=\text{Measure the atomic
system} $ $I= \text{Initialize the atomic system in the } (\ket{1}+
\ket{2})/\sqrt{2} \text{ state}$ and $H$ indicating to ``hold" the
atomic system in the $\ket{1}$ state in order to delay a photon not
involved in any stabilizer check by $T$. In chip layers two and
four, temporally synchronous photons enter a particular chip,
however only one of these photons is routed into the module. These
stages are indicated by $\{.\}^{U,B}$, denoting the port (either U
or B) that is routed through the bypass in that time step.
Additionally, for a given column of chips, there are two switching
sequences, (one denoted by *) which simply alternates down any given column.}
\label{tab:side}
\end{center}
\end{table}
By examining Fig.~\ref{figure:side} there is the potential for photon collisions for chip layers two and four.  However, whenever two photons
enter a chip simultaneously, one interacts with the module while the other is required to bypass the unit.  Hence the notation used in
Tab.~\ref{tab:side}, $\{.\}^{U,B}$, refers to switching the central port $\{.\}$ to the module while switching the photon in the (U)pper or (B)ottom port
to a bypass line.

\section{Resource requirements.}

While we have only illustrated the network for a $5\times 5$ continuously generated lattice, the patterning of photonic chips and the switching
sequence for each unit extends in two dimensions allowing for the continuous generation of an arbitrary large 3D lattice in a modular way [Fig.~\ref{figure:network}].
The total
number of photonic chips required for the preparation of a large cross section of the lattice is easily calculated.
In general, for an $N \times N$ cross section of cells, $4N^2+4N$ photonic chips are required to continually generate the lattice.

We are able to choose the optimal clock cycle of the
computer, $T$, as the fundamental atom/photon interaction time
within each photonic module.  As shown in Tab.~\ref{tab:side}, the
switching sequence for the preparation network allows for a temporal
window of $T$ for the measurement of the atomic system, within each
device.  As estimated in~\cite{Devitt1,Su1}, depending on the system
used, this rate can be approximately 10ns to 1$\mu$s.  If we choose
$T$ to be the optimal operating rate of the photonic module, then
there is the potential that atomic measurement in the preparation
network is too slow.  This can be overcome by the availability of
more photonic chips.  If more chips are available then we 
construct multiple copies of the preparation network with optical
switches placed between each preparation stage.  While the atomic
systems are being measured from one round of parity checks the next
set of incoming photons are switched to a different group of chips.

For a given ratio of the atom/photon interaction time to atomic
measurement time, $T_{\text{atom}}/T_{\text{module}} \geq 1$, photons are routed to multiple
copies of the preparation network.  Therefore the number of additional
photonic modules/chips will increase by a factor of $\Gamma =
\lceil T_{\text{atom}}/T_{\text{module}}\rceil$ to compensate for slow
measurement.  Resource estimates are consequently related to the
total 2D cross section of the lattice and $\Gamma$. Given that the
number of photonic chips required in the preparation network, for a
continuously generated $N\times N$ cross section of unit cells, is
$4N^2+4N$, the number of fabricated photonic
chips required when atomic measurement is slow will be approximately
$\Gamma(4N^2+4N)$.

To put these resource costs in context we can make a quick estimate
of the resources required to build a quantum computer
capable of solving interesting problems.
Let us choose a logical error rate per time
step of the lattice of $10^{-16}$ to be our target error rate, where a logical time step is
the creation and measurement of a single layer of the cluster.  This gives an approximate 
error rate, per logical non-Clifford $R_z(\pi/8)$ rotation of approximately
$O(10^{-11})$ [Appendix~\ref{sec:appB}].    This error rate would
be sufficiently low to enable the factoring of integers several
thousand binary digits long using Shor's algorithm~\cite{Shor1,MI05,Z98}.  Let
us assume that increasing the separation between defects and the
circumference of defects by two cells reduces the logical error rate
per time step by a factor of 100 [Appendix~\ref{sec:appA}].  Given that the current threshold error rate of
the 3D topological cluster state scheme is $6.7\times 10^{-3}$~\cite{Raussendorf3},
albeit in the absence of loss, this assumption is
equivalent to assuming qubits are affected by un-correlated $Z$ errors with a
probability between $10^{-4}$ and $10^{-5}$ per time step.  Correlated errors, which can be 
produced by the photonic chip when preparing the cluster, effect qubits on two separate lattice structures defined by the cluster 
(the primal and dual lattices) which 
are corrected independently and can be treated as such~\cite{Raussendorf6}.  Complete failure of a photonic chip 
is heralded as the eigenvalue conditions for cells prepared by a faulty chip are not satisfied.  Therefore 
the measurement of a small region the $x-y$ plane of the cluster (along the $z$-axis) 
will identify errors in cluster cells at a much higher rate than the rest of the computer, 
indicating chip-failure.  
\begin{figure*}[ht]
\resizebox{120mm}{!}{\includegraphics{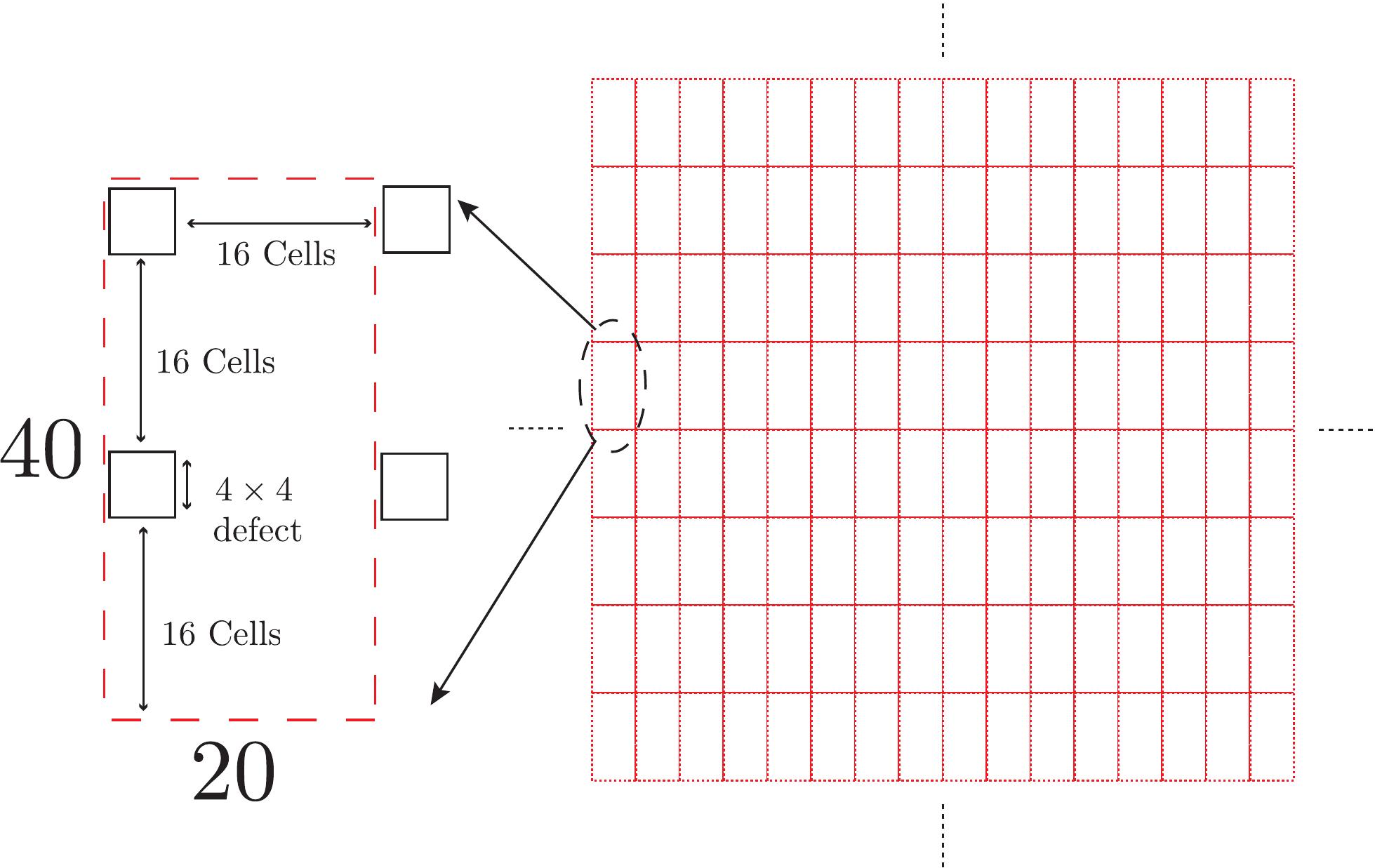}}
\vspace*{-10pt}
\caption{\textbf{Large scale lattice structure required for long term computation via topological cluster states}.  Here we
illustrate a basic concept of the lattice structure that is required for lengthy operations of the topological optical computer
(when examining the lattice along the $z$-axis).
Based on the estimates in the main text, each logical qubit (defined via a pair of defects in the lattice) requires
a $40\times 20$ array of cells.  The layout for each logical qubit is shown on the left, with the larger lattice, comprising thousands (if not millions)
of logical qubits shown on the right.  Each logical qubit requires just over 3000 fabricated photonic chips.  Assuming single qubit error rates in the $10^{-4}$ to $10^{-5}$ range, this lattice would be sufficient 
to permit approximately $10^{11}$ logical non-Clifford $R_z(\pi/8)$ operations.}
\label{figure:perp2}
\end{figure*}

Given the above assumptions, the desired logical error rate per time
step could be achieved with defects measuring $4\times 4$ cells in
cross-section and separated by 16 cells.  
Fig.~\ref{figure:perp2} shows a section of
a semi-infinite lattice of sufficiently well error corrected logical
qubits.  Each logical qubit occupies a cross-sectional region of $40\times 20$ cells.
To prepare such a lattice (for an arbitrary number of time steps and setting $\Gamma = 1$) we require the
fabrication of 3320 integrated photonic chips.

This preliminary estimate illustrates that the resource requirements of
this architecture are promising.  The high fidelity construction of
approximately $3\times 10^{3}$ photonic modules, per logical qubit,
to prepare a lattice that has sufficient topological protection to
perform on the order of $10^{11}$ non-Clifford logical operations is
arguably a less significant challenge than the high fidelity
manufacturing of other proposed quantum computer architectures.
Other proposed systems not only require comparable (if not more)
physical qubits to achieve the same error protection, but depending
on the system they will also require interconnected quantum bus systems
for qubit transport, very non-trivial classical control structures
and most likely the fabrication of the entire computer at once, with
little flexibility to expand the size of the computer as more
resources become available.  

In Appendix~\ref{sec:appB} it is estimated that each $R_z(\pi/8)$ rotation requires approximately 250 logical 
qubits (required for ancilla state distillation in order to enact non-Clifford gates).  It should be noted that the logical qubit requirements 
for non-Clifford gates are identical to all computational models employing state injection, state distillation and teleportation as 
a method for fault-tolerant universal computation.  

\section{Conclusions}

We have presented a detailed architecture for topological cluster state computation in optical systems.  
While this architectural design is not appropriate for all systems, it does 
contain does contain several key elements, 
easing the conceptual design of a large scale computer.  For example: 
\begin{enumerate}
\item The utilization of 
a computational model fundamentally constructed from error correction, rather than implementing error correction codes 
on top of an otherwise independent computational model.
\item  The modular construction of the 
architecture, where architectural expansion is achieved via the addition of comparatively simple elements in 
a regular and known manner.  
\item Utilization of a computational model exhibiting high fault-tolerant thresholds and 
correcting otherwise pathological error channels (such as qubit loss).  
\item Utilization of a measurement based computational model.  As the preparation network is 
designed simply to prepare the quantum resource, programming such a device is a problem of software, not 
hardware.
\end{enumerate}

By constructing this architecture with the
photonic module and photonic chip as the primary operational elements, this design addresses 
many of the significant hurdles limiting the practical scalability of optical quantum computation.  These include
inherent engineering problems associated with probabilistic photon/photon interactions
and the apparent intractability of designing a large scale, programmable system which can be scaled to 
millions of qubits.  

The experimental feasibility of constructing this type of computer is promising.  High fidelity coupling of 
single photons to colour centers in cavities is a significant area of research in the quantum computing 
community and the nano engineering required to produce a high fidelity photonic module is arguably 
much simpler than a full scale atom/cavity based quantum computer which would require the fabrication 
of all data qubits and interconnected transport buses simultaneously.  
An additional benefit is the continuous 
nature of the lattice preparation.  The required number of photonic modules and photonic chips only
depends on the 2D cross sectional size of the lattice.  This is important.  Unlike other computational systems, this 
model does not require us to penalize individual photonic qubits within our physical resource analysis.  
Working under the assumption 
of appropriate on-demand sources, resource costs become a function of the total number of 
photonic chips (and single photon sources), rather that the total number of {\em actual} 
qubits required for a large 3D cluster.  
This highlights the advantages of having photons as disposable computational resources and we hope 
recasts the question of quantum resources into how many active quantum components are 
required to construct a large scale computer.


\section{Acknowledgments}
The authors thank C.-H. Su, Z. W. E. Evans and C. D. Hill for helpful
discussions. The authors acknowledge the support of MEXT, JST, the
EU project QAP, the Australian Research Council (Centre of
Excellence Scheme and Fellowships DP0880466, DP0770715), and the
Australian Government, the US National Security Agency, Army
Research Office under Contract No. W911NF-05-1-0284.

\appendix
\section{Error Scaling of the Topological Lattice}
\label{sec:appA}

This appendix briefly reviews how to use the three-dimensional cluster state
that is prepared by the network. In particular we outline techniques
for error correction and universal computation and explain how
logical failure rates are suppressed and how resource
requirements are calculated.

To extract the error syndrome, one measures each face qubit in every
unit cell (see Fig.~\ref{figure:unit}) in the $X$ basis. In the absence of $Z$
errors on these qubits, the parity of these six measurements will be
even. A single $Z$ error on one face qubit, or an error during
measurement, will flip the parity that is recorded from even to odd.
When a chain of more than one error occurs, only cells at the end
points of the chain will record odd parity. Note that no information
about the path of any error chains is obtained during error
correction and that error chains can terminate on boundaries of the
cluster.

Identifying the minimal set of $Z$ errors that is consistent with
the syndrome proceeds by representing each odd parity unit cell as a
node in a completely connected, weighted graph. The weight of the edge
between any two nodes is the minimum number of errors that could
connect the two cells which they represent. For each node in the
graph a partner node is added which represents a boundary where an
error chain could have terminated. The weight of the edge connecting
each node to its partner node is the minimum number of errors that
could connect the cell to the boundary. Partner nodes also form a
fully connected graph with the weight between any two nodes equal to
zero. The entire graph is then solved for the minimum weight pairing
of the nodes.Ê Each pair in the solution represents a $Z$ error or
chain of $Z$ errors which can be corrected for with classical
post-processing.

We note that only $Z$ error correction is required to correct $Z$,
$X$, measurement, initialization errors and photon loss. The
measurements required for syndrome extraction are made in the $X$
basis and so are not affected by $X$ errors. Errors during
preparation of the cluster state are equivalent to $Z$ errors on the
prepared state. Photon loss could be treated as a random measurement 
result, equivalent to a measurement error or alternatively by calculating the 
parity of a closed surface of face qubits enclosing the lost photon or photons.  This 
latter method is preferable given recent results establishing a high tolerance to 
heralded loss events~\cite{Stace1}.

A single logical qubit is associated with a pair of defects in the
lattice. A defect is a connected volume of the lattice in which all
qubits have been measured in the $Z$ basis. Two types of defects can
be made - primal and dual. Primal defects are regions where one or
more unit cells (see Fig.~\ref{figure:unit}) have been measured. Dual defects are
defined identically in the dual space of the lattice, where a center
of a dual unit cell corresponds to a vertex of a primal unit cell.
Valid operations on a single logical qubit include joining the
pair of defects via a chain of physical $Z$ operations and encircling
a single defect via a chain of physical $Z$ operations. Technically, these 
logical operations have additional $X$ operators associated with them to 
ensure anticommutation~\cite{Fowler2}, however these do not 
affect $X$ basis measurements and thus can be ignored in the current discussion. 
Note that logical $X$ and $Z$ operations can
be realized by simply tracing their  effect through the circuit until
logical measurement and then appropriately adjusting the measurement
outcome, as opposed to performing gated operations directly on the
physical qubits within the three-dimensional lattice. A
controlled-\textsc{not} interaction between a dual and primal qubit
can be effected by braiding one of the dual defects around one of
the primal defects. Braiding is performed by changing the shape of
the region of measurements defining the location of a defect. Logical CNOT, 
combined with initialization, readout, state
injection, and state distillation allow universal quantum
computation as detailed in Ref~\cite{Fowler2}.

Protection against logical errors is achieved by increasing the
separation between defects and the circumference of defects. A
logical error occurs when the error correction routine causes the
application of a set of corrections that make or complete a chain of
physical operations forming an erroneous logical operation. Just
considering error chains joining defects, it is clear that the
number of physical errors required to make such a chain increases
linearly with the defect separation. The probability of forming such
a chain therefore decreases exponentially with separation. As the
distance of a code can be defined as the weight of its minimal
weight logical operator, increasing the separation between defects
by two cells increases the code distance by two. Increasing the
circumference of defects has an analogous effect on the probability
of logical errors in the conjugate basis.

The efficacy of error correction depends on the amount by which the
physical error rate, $p$, is below some threshold error rate. As the
distance of a code is increased from $d$ to $d+2$, the encoded error
rate is transformed from $n(d)p^{(d+1)/2}$ to $n(d+2)p^{(d+1)/2+1}$,
where higher order terms are neglected and $n(d)$ and $n(d+2)$ are
constants related to the code and the circuits required to extract
the syndrome. Increasing the distance of the code will lower the
encoded error rate only if $p<n(d)/n(d+2)$. If $p$ is a factor of
$x$ below this threshold ($p = n(d)/xn(d+2)$), 
then the encoded error rate is reduced by a factor 
$n(d)p^{(d+1)/2}/n(d+2)p^{(d+1)/2+1} = n(d)/pn(d+2) = x$.  
In general, provided that the quantity
$n(d)/n(d+2)$ is constant for all $d$, increasing the distance of a
code by two will result in a reduction in the encoded error rate by
a factor equal to the difference between $p$ and the threshold error
rate. In our analysis of topological error correction we assume that
$n(d)/n(d+2)$ is constant for all $d$.

The threshold for the topological error correction code described
above is $6.7\times 10^{-3}$, where qubit loss is neglected and
where all other errors are assumed to be equally likely to
occur~\cite{Raussendorf5}. If we assume that the physical error rate
is two orders of magnitude below this threshold, equal to $6.7\times
10^{-5}$, then increasing the separation of defects and the
circumference of defects by two cells reduces the encoded error by a
factor of 100. If our target encoded error rate, per time step, is $10^{-16}$ we
require a minimum code distance of 17, which corresponds to a defect
separation of 16 unit cells, or alternatively that we can correct
error chains up to 8 errors long. Similarly the perimeter of the
cells should also be 16, and hence we require defects that are
$4\times 4$ cells in cross-section. This directly leads to the
resource estimates per logical qubit given in the body of the paper.

\section{Error rate on non-Clifford logical operations}
\label{sec:appB}

In the main body of the text, two logical error rates are presented.  The first is the effective logical error 
rate per single layer of the cluster along the direction of simulated time [Appendix~\ref{sec:appA}]
and the second is the effective error rate per non-Clifford rotation $R_z(\pi/8)$.  We present these 
two values separately as the optimization of the measurement sequence for applying non-Clifford 
gates is still incomplete.  This calculation instead provides a rough order of 
magnitude estimate of the effective non-Clifford error rate given the topological 
protection afforded by the lattice specified in the 
body of the paper.  

Within the topological model, only a subset of universal gates can be applied directly to the 
lattice, these include initialization and measurement in the $X$ and $Z$ basis, single qubit 
$X$ and $Z$ operations (although these are realized by tracing their effect through the circuit until
logical measurement and then appropriately adjusting the results) and braided CNOT gates.  
Completing the universal set is achieved through state injection, magic state distillation and 
teleportation protocols~\cite{Raussendorf5,BK05}.  

As non-Clifford rotations require the implementation of state distillation protocols, 
the effective logical gate rate is then dependent on the volume of cluster used 
to inject and distill a sufficiently high fidelity ancilla state for use in the teleportation 
protocol.

\begin{figure}[ht]
\resizebox{60mm}{!}{\includegraphics{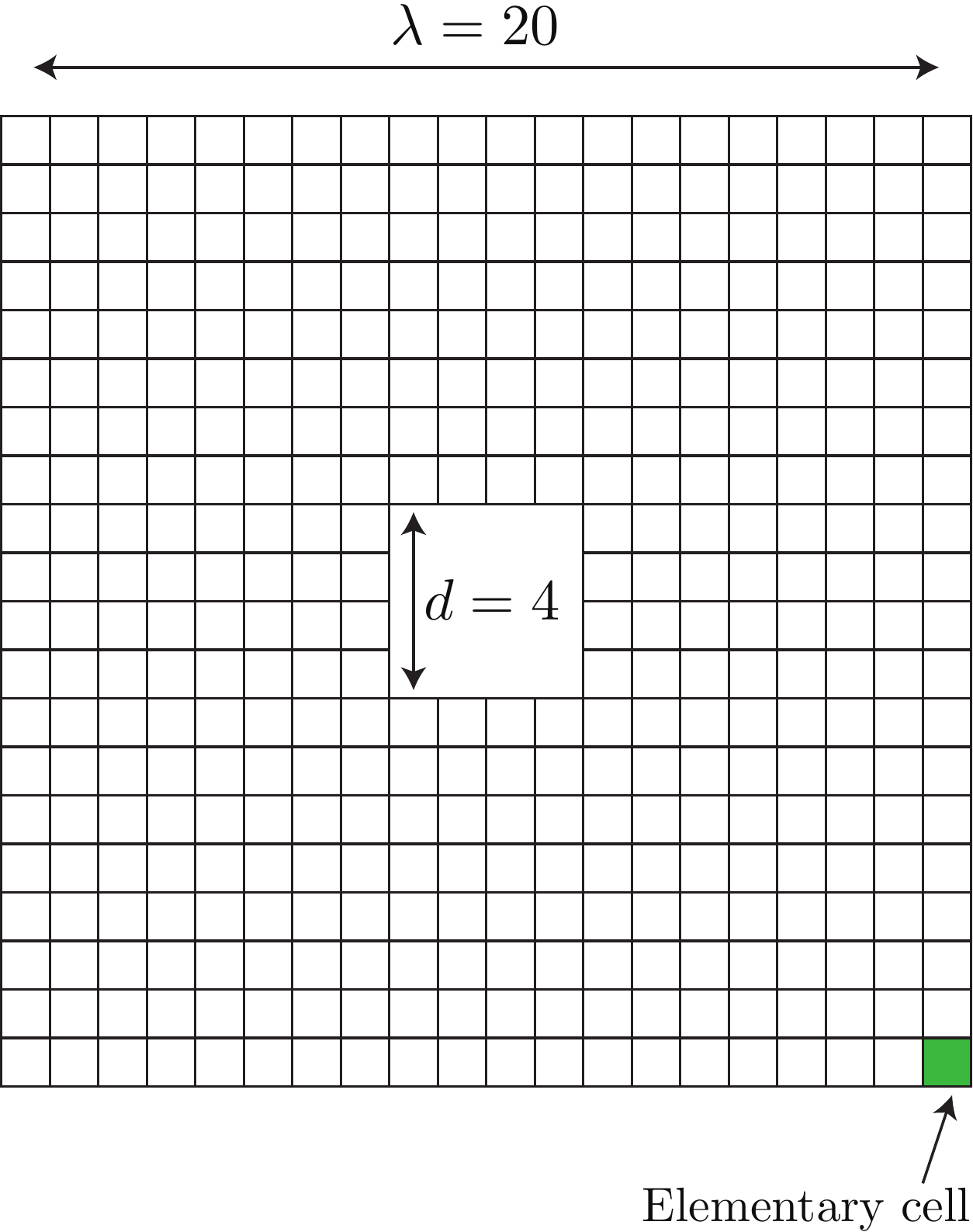}}
\vspace*{-10pt}
\caption{\textbf{(From Ref.~\cite{Raussendorf5}) Cross section of a re-scaled logical cell providing topological protection.} 
To achieve sufficient fault-tolerant protection, each elementary cell of the cluster [Fig.~\ref{figure:unit}] is 
rescaled to a larger logical cell, where a central defect has sufficient circumference and separation 
from other defects to be highly protected against error.  Shown here is the 2D cross section of 
a rescaled logical cell in terms of the lattice parameters utilized in the main text [Fig.~\ref{figure:perp2}].}
\label{figure:rescale}
\end{figure}

\begin{figure}[ht]
\resizebox{80mm}{!}{\includegraphics{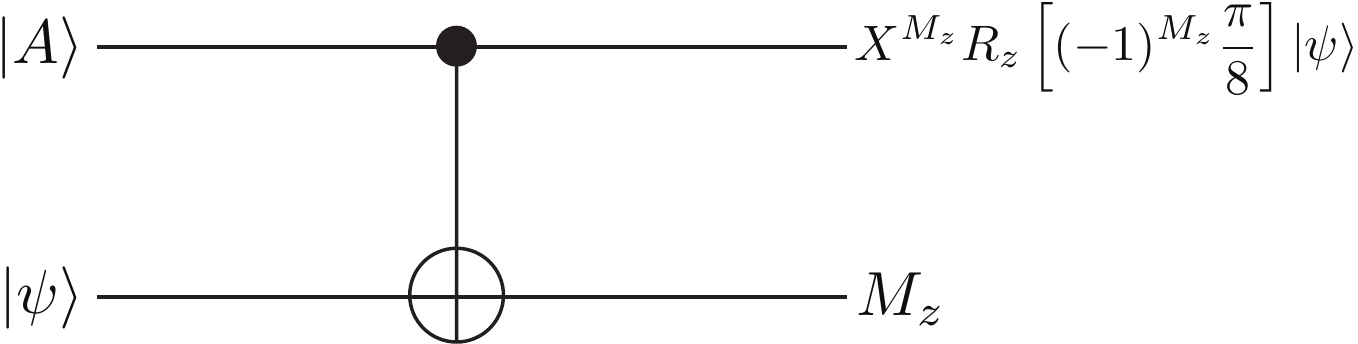}}
\caption{\textbf{Standard Telegate circuit to implement single qubit $T$ rotation.} 
Given the availability of the ancilla state $\ket{A}$, this circuit allows for the teleported implementation 
of the $T$ gate on an unknown state, $\ket{\psi}$.  The measurement result, $M_z$, determines if 
the rotation $T$ or $T^{\dagger} \equiv R_z(-\pi/8)$ is applied.  
As this measurement result is random, with a $T$ gate 
applied 50\% of the time, if $T^{\dagger}$ is applied 
a second teleportation circuit is run to implement the gate 
$P$ ($PT^{\dagger} = T$).}
\label{figure:teleport}
\end{figure}

In order to estimate the failure rate of a non-Clifford $R_z(\pi/8)$ operation, we examine the volume 
required to distill high fidelity 
ancilla states from low fidelity injected sources and to perform the required teleportation circuits 
to implement the $R_z(\pi/8)$ gate.
Tab.~\ref{tab:gate} from Ref.~\cite{Raussendorf5} specifies the volume of the 3D cluster that is required to implement 
five specific gate operations within the TCQC model, namely the CNOT, state distillation circuits for 
the singular qubit states,
\begin{equation}
\ket{Y} = \frac{1}{\sqrt{2}}\left ( \ket{0}+i\ket{1}\right ), \quad  \ket{A} = \frac{1}{\sqrt{2}}\left ( \ket{0}+e^{i\frac{\pi}{4}}\ket{1}\right )
\label{eq:ancilla}
\end{equation}
and teleportation circuits to implement the single qubit gates,
\begin{equation}
P = \begin{pmatrix} 1 & 0 \\ 0 & i \end{pmatrix}, \quad R_z(\pi/8) \equiv T = \begin{pmatrix} 1 & 0 \\ 0 & e^{i\frac{\pi}{4}} 
\end{pmatrix}
\end{equation}
\begin{table}[ht!]
\begin{center}
\vspace*{4pt}
\begin{tabular}{c|c}
Gate & Volume \\
\hline
CNOT & $V_2=12$ (16) \\
Telegate, $T$ & $V_{1,z} =2$ \\
Telegate, $P$ & $V_{1,x} = 4$ \\
$\ket{A}$-distillation & $V_A=336$ (168) \\
$\ket{Y}$-distillation & $V_Y=120$ (60)\\
\end{tabular}
\vspace*{4pt} \caption{Total volume utilized in the 3D cluster to perform logical gates. In brackets 
are revised volume estimates~\cite{Austin} that are currently unpublished. Our calculations will 
assume the original volume estimates from Ref.~\cite{Raussendorf5}, overestimating the total volume 
consumed for the gate $R_z(\pi/8)$.}
\label{tab:gate}
\end{center}
\end{table}
The volume estimates are given in terms of a scaled logical cell, where 
defects are now defined such that they have a sufficiently large circumference and separation to 
suppress the probability of logical failure, Fig.~\ref{figure:rescale} (from 
Ref.~\cite{Raussendorf5}) illustrates.  The cluster volumes quoted in 
Tab.~\ref{tab:gate} give the total number of logical cells required to implement specific gates.    
Fig.~\ref{figure:rescale} shows defects of the same dimensions 
to Fig.~\ref{figure:perp2}, but translated along the $x-y$ plane 
of the lattice.  A re-scaled logical cell is now a volume of $\lambda^3 = 20^3$ with a cylindrical defect of 
volume $d^2\lambda = 16\times 20$ passing through its center.  
To implement the non-Clifford gate, $T$, injected low fidelity states [Eq.~\ref{eq:ancilla}] need to be purified using 
magic state distillation~\cite{BK05} before teleportation circuits can be used to enact the gate.  Shown in 
Fig.~\ref{figure:teleport} is the quantum circuit required to perform a single qubit $T$ gate on 
an arbitrary state $\ket{\psi}$ given an appropriate ancilla $\ket{A}$.  The measurement 
result of the ancilla qubit after the CNOT determines if the gate $T$ or $T^{\dagger}$ is 
applied.  If the gate $T^{\dagger}$ is applied, then the further application of a single 
qubit $P$ gate transforms the rotation from $T^{\dagger}$ to $PT^{\dagger} = T$.  
As single qubit $P$ gates also require distilled ancilla and teleportation protocols and the 
application of the teleported gate, $T$, occurs with a probability of 0.5, every two $T$ gates within 
the quantum circuit will, on average, require the application of one $P$ gate.  Hence not only 
do we need to distill one $\ket{A}$ state, but we also need to distill ``half" a $\ket{Y}$ state.  

Assuming that the residual error on all injected qubits within the lattice is equal to our assumed 
operational error rate of the computer $p = 6.7\times 10^{-5}$, two concatenated levels of 
state distillation are required.  To leading order in $p$, the recursion relations and success 
probabilities ($P_{A,Y}$) for $\ket{A}$ and $\ket{Y}$ state distillation, in the limit of negligible 
topological error, are given by,
\begin{equation}
\begin{aligned}
p_{l+1}^A = 35(p_l^A)^3, \quad p_{l+1}^Y = 7(p_l^Y)^3, \\
P_{l+1}^A = 1-15p_{l}^A, \quad P_{l+1}^Y = 1-7p_{l}^Y, \\
\end{aligned}
\end{equation}
for concatenation level $l$~\cite{Raussendorf6}.  For $p_0^A = 6.7\times 10^{-5}$ 
and $p_0^Y =6.7\times 10^{-5}$, the residual error after 
two rounds of distillation is $p_2^A \approx O(10^{-32})$ and $p_2^Y \approx O(10^{-35})$ 
and the probability of 
obtaining an incorrect syndrome result for each level is $[P_1^A \approx O(10^{-3})$, 
$P_2^A \approx O(10^{-10})]$ and [$P_1^Y \approx O(10^{-4})$, 
$P_2^Y \approx O(10^{-11})$].  

In total, $15/(1-P_1^A) + 1/(1-P_2^A) \approx 16$ 
distillation circuits are required for two levels of $\ket{A}$ state distillation 
and $7/(1-P_1^Y)+1/(1-P_2^Y)\approx 8$ are required for two levels of $\ket{Y}$ state distillation. 
Run in parallel, this requires a total volume of $16V_A+(1/2)\times 8V_Y = 5856 $ logical cells, where the factor of $1/2$ accounts for the probabilistic implementation of the $P$ gate.  
The teleportation circuit is then applied requiring a logical volume of $V_{1,z}+(1/2)V_{1,x} = 4$ cells.
Given that the scaling factor for the computer detailed in Fig.~\ref{figure:perp2} is $\lambda = 20$, 
the total number of elementary cells, {\em in the direction of simulated time}, is
$\Omega = \lambda(5865+4) \approx 1.1\times 10^5$ and the failure rate of the logical $T$ gate is 
approximately,
\begin{equation}
1-(1-10^{-16})^\Omega \approx O(10^{-11})
\end{equation}
Hence, this basic estimate illustrates that the logical non-Clifford failure rate for the 
topological computer 
is reduced by approximately five orders of magnitude from the logical failure rate per temporal layer.  
Note that this value for the logical gate 
assumes no specific optimization of non-Clifford 
group gates and represents a conservative estimate on the expected logical gate fidelity.  In addition to the error rate of the logical gate, a single 
application of the $T$ gate consumes, on average, $15^2+7^2/2 \approx 250$ logical qubits in the lattice.  However, the qubit resources at the logical 
level are equivalent for {\em all} computational models employing state distillation and teleportation of $P$ and $T$ gates to achieve universality.

\clearpage
\newpage
\onecolumngrid


\begin{thebibliography}{1}

\bibitem{NPT99}
Y. Nakamura, Yu. A. Pashkin and J.S. Tsai.
\newblock {\em Nature (London)} \textbf{398}, 786 (1999).

\bibitem{CNHM03}
I. Chiorescu, Y. Nakamura, C.J.P.M. Harmans and J.E. Mooij,
\newblock {\em Science} \textbf{299}, 1869 (2003).

\bibitem{YPANT03}
T. Yamamoto, Yu. A. Pashkin, O. Astafiev, Y. Nakamura and J.S. Tsai
\newblock {\em Nature (London)} \textbf{425}, 941 (2003).

\bibitem{CLW04}
J. Chiaverini, D. Leibfried, T. Schaetz, M.D. Barrett, R.B. Blakestad, J. Britton, W.M. Itano, J.D. Jost, E. Knill, C. Langer, R. Ozeri and D.J. Wineland.
\newblock {\em Nature (London)} \textbf{432}, 602 (2004).

\bibitem{HHB05}
H. H\"{a}ffner, W. H\"{a}nsel, C.F. Roos, J. Benhelm, D. Chek-al-kar, M. Chwalla, T. K\"{o}rber, U.D. Rapol, M. Riebe, P.O. Schmidt, C. Becher, O. G\"{u}hne, W. D\"{u}r and R. Blatt.
\newblock {\em Nature (London)} \textbf{438}, 643 (2005).

\bibitem{GHW05}
J. Gorman, D.G. Hasko and D.A. Williams.
\newblock {\em Phys. Rev. A.} \textbf{95}, 090502 (2005).

\bibitem{G06}
T. Gaebel, M. Domhan, I. Popa, C. Wittmann, P. Neumann, F. Jelezko, J.R. Rabeau, N. Stavrias, A.D. Greentree, S. Prawer, J. Meijer, J. Twamley, P.R. Hemmer and J. Wrachtrup.
\newblock{ Nature (Physics)} \textbf{2}, 408 (2006).

\bibitem{H06}
R. Hanson, F.M. Mendoza, R.J. Epstein, and D.D. Awschalom.
\newblock {\em Phys. Rev. Lett.} \textbf{97}, 087601 (2006).

\bibitem{G07}
M.V. Gurudev Dutt, L. Childress, L. Jiang, E. Togan, J. Maze, F. Jelezko, A.S. Zibrov, P.R. Hemmer, and M.D. Lukin.
\newblock {\em Science} \textbf{316}, 1312 (2007).

\bibitem{O'Brien2}
J.L.~OÕBrien, G.J.~Pryde, A.G.~White, T.C.~Ralph, and D.~Branning.
\newblock {\em Nature (London)} \textbf{426}, 264 (2003).

\bibitem{Kielpinski1}
D.~Kielpinski, C.~Monroe, and D.~J.~Wineland.
\newblock {\em Nature} \textbf{417}, 709 (2002).

\bibitem{Taylor1}
J.~M.~Taylor, H.-A.~Engel, W.~D\"ur, A.~Yacoby, C.~M.~Marcus, P.~Zoller, and M.~D.~Lukin.
\newblock {\em Nature Phys.} \textbf{1}, 177 (2005).

\bibitem{Hollenberg1}
L.~C.~L.~Hollenberg, A.~D.~Greentree, A.~G.~Fowler, and C.~J.~Wellard.
\newblock {\em Phys. Rev. B} \textbf{74}, 045311 (2006).

\bibitem{Fowler3}
A.~G. Fowler, W.~F.~Thompson, Z.~Yan, A.~M. Stephens, B.~L.~T.~Plourde, and F.~K.~Wilhelm.
\newblock {\em Phys. Rev. B} \textbf{76}, 174507 (2007).

\bibitem{Knill1}
E.~Knill, R.~Laflamme, and G.~J.~Milburn.
\newblock {\em Nature} \textbf{409}, 46 (2001).

\bibitem{Kok1}
P.~Kok, W.~J.~Munro, K.~Nemoto, T.~C.~Ralph, J.~P.~Dowling, and G.~J.~Milburn
\newblock {\em Rev. Mod. Phys.} \textbf{79}, 135 (2007).

\bibitem{Raussendorf4}
R. Raussendorf and J. Harrington,
\newblock {\em Phys. Rev. Lett.} \textbf{98}, 190504 (2007).

\bibitem{Raussendorf5}
R. Raussendorf, J. Harrington and K. Goyal,
\newblock {\em New J. Phys.} \textbf{9}, 199 (2007).

\bibitem{Fowler2}
A.G. Fowler and K. Goyal,
\newblock {\em arXiv:0805.3202} (2008).

\bibitem{Raussendorf1}
R. Raussendorf, H.J. Briegel,
\newblock {\em Phys. Rev. Lett.} {\bf 86}, 5188 (2001).

\bibitem{NC00}
M.A. Nielsen and I.L. Chuang,
\newblock {\em Quantum Information and Computation.} Cambridge University Press (2000).


\bibitem{Nayak1}
C. Nayak, S.H. Simon, A. Stern, M. Freedman and S. Das Sarma,
\newblock {\em arXiv:0707.1889} (2007).

\bibitem{BK01}
S.B. Bravyi and A.Y. Kitaev,
\newblock {Quant. Computers and Computing} \textbf{2}, 43 (2001).

\bibitem{DKLP02}
E. Dennis, A.Y. Kitaev, A. Landahl and J. Preskull
\newblock {J. Math. Phys.} \textbf{43}, 4452 (2002).

\bibitem{Fowler1}
A.G. Fowler, A.M. Stephens and P. Groszkowski,
\newblock {\em arXiv:0803.0272} (2008).


\bibitem{Dawson2}
C.M. Dawson, H.L. Haselgrove and M.A. Nielsen,
\newblock {\em Phys. Rev. A}, \textbf{73}, 052306 (2006).

\bibitem{Ralph1}
T.C. Ralph, A.J.F. Hayes and A. Gilchrist,
\newblock {\em Phys. Rev. Lett.} \textbf{95}, 100501 (2005).


\bibitem{Varnava1}
M. Varnava, D.E. Browne and T. Rudolph,
\newblock{\em Phys. Rev. Lett.}, \textbf{100}, 160502 (2008).






%
%



\bibitem{Beenakker4}
C.W.J. Beenakker, D.P. DiVincenzo, C. Emary, and M. Kindermann,
{\em Phys. Rev. Lett.}  \textbf{93}, 020501 (2004);

\bibitem{Nemoto4}
K. Nemoto and W.J. Munro,
{\em Phys. Rev. Lett.}  \textbf{93}, 250502 (2004).

\bibitem{Spiller6}
T.P. Spiller, K. Nemoto, S.L. Braunstein, W.J. Munro, P. van Loock, G.J. Milburn,
{\em New J. Phys.} \textbf{8}, 30 (2006).

\bibitem{Domokos1}
P. Domokos, J.M. Raimond, M. Brune and S. Haroche.
\newblock {\em Phys. Rev. A.} \textbf{52}, 3554 (1995).

\bibitem{Duan1}
L.-M. Duan and H.J. Kimble.
\newblock {\em Phys. Rev. Lett.} \textbf{92}, 127902 (2004).

\bibitem{Devitt1}
S.J. Devitt, A.D. Greentree, R. Ionicioiu, J.L. O'Brien, W.J. Munro and L.C.L. Hollenberg
\newblock {\em Phys. Rev. A.} \textbf{76}, 052312 (2007).

\bibitem{Stephens1}
A.M. Stephens, Z.W.E. Evans, S.J. Devitt, A.G. Fowler, A.D. Greentree, W.J. Munro, J.L. O'Brien,
K. Nemoto and L.C.L. Hollenberg,
\newblock {\em Phys. Rev. A.}, \textbf{78}, 032318 (2008).

\bibitem{Su1}
C-H. Su, A.D. Greentree, W.J. Munro, K. Nemoto, L.C.L. Hollenberg
\newblock {\em Phys. Rev. A.}, \textbf{78}, 062336 (2008).

\bibitem{Yang1}
W.L. Yang, H. Wei, F. Zhou and M. Feng
\newblock {\em J. Phys. B: At. Mol. Opt. Phys.} \textbf{42}, 055503 (2009)

\bibitem{E1}
D. Englund, A. Faron, I. Fushman, N. Stoltz, P. Petroff, and J. Vuckovic
\newblock {\em Nature (London)}, \textbf{450}, 857 (2007).

\bibitem{F1}
I. Fushman, D. Englund, A. Faron, N. Stoltz, P. Petroff, and J. Vuckovic
\newblock {\em Science}, \textbf{320}, 769 (2008).

\bibitem{F2}
A. Faron, I. Fushman, D. Englund, N. Stoltz, P. Petroff, and J. Vuckovic
\newblock {\em Optics Express}, \textbf{16}, 12154 (2008).

\bibitem{P1}
A. Politi, M.J. Cryan, J.G. Rarity, S. Yu and J.L. O'Brien
\newblock {\em Science} \textbf{320}, 646 (2008).

\bibitem{C1}
A.S. Clark, J. Fulconis, J.G. Rarity, W.J. Wadsworth and J.L. O'Brien
\newblock {\em arXiv:0802.1676}, (2008).

\bibitem{M1}
G.D. Marshall, A. Politi, J.C.F. Matthews, P. Dekker, M. Ams, M.J. Withford and J.L. O'Brien
\newblock {\em arXiv:0902.4357}, (2009).

\bibitem{Gottesman1}
D. Gottesman,
Ph.D Thesis (Caltech),
\newblock {\em quant-ph/9705052}, (1997).



\bibitem{Shor1}
P.W. Shor,
\newblock{\em SIAM J. of Computing}, \textbf{26}, 1484 (1997).

\bibitem{MI05}
R. Van Meter and K.M. Itoh
\newblock {\em Phys. Rev. A.} \textbf{71}, 052320 (2005).

\bibitem{Z98}
C. Zalka
\newblock {\em quant-ph/9806084} (1998).

\bibitem{Raussendorf3}
R. Raussendorf, J. Harrington and K. Goyal,
\newblock{\em Ann. Phys.} \textbf{321}, 2242 (2006).

\bibitem{Nemoto1}
K. Nemoto and W.J. Munro,
\newblock{\em Phys. Lett. A.} \textbf{344}, 104 (2005).

\bibitem{Stace1}
T.M. Stace, S.D. Barrett and A.C. Doherty,
\newblock {\em arXiv:0904.3556} (2009)

\bibitem{BK05}
S. Bravyi and A. Kitaev
\newblock {\em Phys. Rev. A.} \textbf{71}, 022316 (2005).

\bibitem{Austin}
A.G. Fowler {\em et. al.} in Preparation (2009).

\bibitem{Raussendorf6}
R. Raussendorf, J. Harrington and K. Goyal
\newblock {\em Ann. Phys.} \textbf{321}, 2242 (2006)

\end{thebibliography}
\end{document}